\documentclass[10pt,conference]{IEEEtran}
\usepackage{multirow}
\usepackage{tabularx} % specify table width
\usepackage[small, tableposition=top]{caption}  
\usepackage{cuted}  
\usepackage{algorithm}
\usepackage{algpseudocode}
\usepackage{multicol, blindtext}
\usepackage{adjustbox}
\usepackage{lipsum,booktabs}
\usepackage{color}
\usepackage{verbatim}
\usepackage{amssymb}
\usepackage{subfigure}
\usepackage{graphicx}
\usepackage{hyperref}
\usepackage{float}
\usepackage{subcaption}
\usepackage{makecell}

\ifCLASSOPTIONcompsoc

  \usepackage[nocompress]{cite}
\else
  \usepackage{cite}
\fi

\ifCLASSINFOpdf
\else
 
\fi
\makeatletter % added <<<<<<<<<<<<<<<<<<<
\def\subsubsection{\@startsection{subsubsection}{3}{}{2.5ex plus 0.1ex minus 0.1ex}%
    {2ex plus .5ex minus 0ex}{\normalfont\normalsize\bfseries}}%
\makeatother

\usepackage{longtable}

\begin{document}

\title{Multi-stage Dynamic Selection for Cross-Project Defect Prediction\\

\thanks{}
}

\author{
\IEEEauthorblockN{Juscimara G. Avelino$^{\dagger}$,
Juscelino S. A. Junior$^{\dagger}$,
George D. C. Cavalcanti$^{\dagger}$,
Rafael M. O. Cruz$^{\ddagger}$}

\IEEEauthorblockA{$^{\dagger}$Centro de Informática, Universidade Federal de Pernambuco, Recife, Brazil}

\IEEEauthorblockA{$^{\ddagger}$École de Technologie Supérieure, University of Quebec, Montreal, Canada}

\IEEEauthorblockA{\texttt{\{jga2, jsaj, gdcc\}@cin.ufpe.br, rafael.menelau-cruz@etsmtl.ca}}
}
% \author{\IEEEauthorblockN{Juscimara G. Avelino}
% \IEEEauthorblockA{Centro de Informática (CIn)\\
% Universidade Federal de Pernambuco\\
% Recife - PE, Brasil\\
% Email: jga2@cin.ufpe.br}
% \and
% \IEEEauthorblockN{Juscelino S. A. Júnior}
% \IEEEauthorblockA{Centro de Informática (CIn)\\
% Universidade Federal de Pernambuco\\
% Recife - PE, Brasil\\
% Email: jsaj@cin.ufpe.br}
% \and
% \IEEEauthorblockN{George D. C. Cavalcanti}
% \IEEEauthorblockA{Centro de Informática (CIn)\\
% Universidade Federal de Pernambuco\\
% Recife - PE, Brasil\\
% Email: gdcc@cin.ufpe.br}
% \and
% \IEEEauthorblockN{Rafael M. O. Cruz}
% \IEEEauthorblockA{École de Technologie Supérieure\\University of Quebec\\ Montreal - Quebec, Canada\\
% Email: rafael.menelau-cruz@etsmtl.ca}} 

\maketitle

\begin{abstract}
%Defect prediction is a crucial task in software engineering, which aims to identify and allocate resources to predict potentially defective software modules. However, building such prediction models requires large amounts of labeled data which is impractical as some companies or projects may lack historical data. 
%which have proven to overcome single classifiers. 

Cross-Project Defect Prediction (CPDP) involves building models using data from external projects, called training projects, to predict modules from the target project. However, traditional CPDP methods suffer from the distribution shift between training and target projects that affects the model's performance. This paper proposes a novel CPDP framework that addresses this issue by proposing a two-stage multiple classifier system (MCS) selection scheme: one working at the project level and another at the module level. In the first stage, the framework evaluates multiple possible MCS configurations to find one that covers and generalizes well across multiple training projects. Consequently, the proposal is likely to obtain a diverse set of classifiers, each specialized in tackling software modules with distinct characteristics. The second selection stage operates at test time, selecting the most competent classifiers to predict each new module in the target project. Unlike previous approaches that apply the same classifiers to the entire target project, the proposed framework performs module-level model selection. This way, the system is more robust to changes in distributions between training and target projects because the selected set of classifiers is module-dependent. Our experimental results using 82 projects from four different CPDP benchmark datasets demonstrate that the proposed approach outperforms the state-of-the-art CPDP methods in most scenarios. The code, dataset, and further details about the proposed method are publicly available at \url{https://github.com/jsaj/Multi_DES}.

\end{abstract}

\IEEEpeerreviewmaketitle

\section{Introduction}

Software reliability is a key factor for ensuring the quality of software products. Software defects are programming problems that may arise from issues in the source code or requirements, negatively affecting software quality and reliability \cite{rawat2012software}. Software Defect Prediction (SDP) aims to support resource allocation by identifying defect-prone software modules in advance \cite{ni2020revisiting}. To this end, the research community has proposed several machine learning–based approaches that leverage data stored in software repositories to predict defect proneness or estimate the likelihood of future failures \cite{malhotra2015systematic}.

Although a wide range of defect prediction methods has been proposed, building accurate models requires sufficient historical defect data \cite{hall2011systematic}. In practice, however, many software projects lack such data, especially newly created projects or those that do not maintain detailed defect histories \cite{kitchenham2007cross}. As a result, the limited availability of training data remains a major challenge in software defect prediction.

To address this issue, Cross-Project Defect Prediction (CPDP) has been proposed as an alternative, in which models are trained on data from external projects (source projects) and applied to a target project \cite{watanabe2008adapting,herbold2013training}. CPDP is particularly valuable when the target project has insufficient historical data and has therefore become an important subtopic in defect prediction research \cite{herbold2017comparative}. Despite its potential, CPDP methods face major challenges due to distribution shifts between source and target projects and the limited availability of supervised or unsupervised data from the target project. As such, empirical studies have shown that many state-of-the-art CPDP approaches are not significantly better than trivial predictors \cite{herbold2017comparative}.

Several strategies have been proposed to mitigate distribution shifts in CPDP, including data transformation, source project selection, and ensemble-based weighting schemes \cite{cruz2009towards, liu2019two, panichella2014cross}. However, distribution shifts can still significantly degrade generalization performance \cite{cieslak2009framework}, especially in CPDP scenarios where projects differ in coding standards, development practices, and data collection procedures \cite{rawat2012software}. Most existing approaches tackle this problem globally by modeling entire project distributions and often rely on target project information to estimate shifts or learn transformations. We argue that this global perspective is limited, as CPDP is better viewed as a dynamic problem composed of multiple local regions with distinct characteristics, motivating adaptive techniques that adjust their behavior at prediction time.

In this work, we propose Multi-stage Dynamic Ensemble Selection for Cross-Project Defect Prediction (Multi-DES), a CPDP framework based on a two-stage selection scheme designed to improve robustness under distribution shifts.

Multi-DES operates at two levels. At the project level, it selects the dynamic selection technique, base classifier, and pool size that best generalize across a set of training projects using an Aggregate Rank Minimization (ARM) scheme, a multi-criteria ranking strategy that considers multiple performance metrics \cite{herbold2017comparative}. At the instance level, the selected configuration dynamically identifies the most competent classifiers for each software module in the target project. By relying exclusively on training projects and performing instance-based selection at inference time, Multi-DES can be applied to entirely new projects with very limited data, including scenarios with only a single software module.

The second selection stage relies on Dynamic Selection (DS) techniques \cite{cruz2018dynamic}. DS methods assume that different classifiers act as experts in different regions of the feature space and dynamically select the most competent classifier for each instance. Their ability to adapt the system topology at inference time makes them a promising alternative for handling distribution shifts in heterogeneous scenarios \cite{jiao2022dynamic,zyblewski2021preprocessed}.

Our experimental results show that the proposed Aggregate Rank Minimization (ARM) selection mechanism leads to better generalization than single-metric selection strategies. Multi-DES achieves superior or equivalent performance compared to state-of-the-art CPDP methods in most scenarios while relying exclusively on source project data. 

% The data and source code used in this study will be made publicly available upon acceptance of the paper.

The main contributions of this work are summarized as follows:
\begin{itemize}
    \item A multi-stage dynamic selection framework for CPDP, named Multi-DES, that does not rely on any information from the target project distribution.
    
    \item An Aggregate Rank Minimization (ARM)–based configuration selection strategy that improves generalization compared to traditional single-metric approaches.

    \item A comprehensive empirical evaluation on four CPDP benchmark datasets comprising 82 software projects.
\end{itemize}

\section{Related work}
\label{sec:relatedwork}

This section reviews the main approaches proposed for Cross-Project Defect Prediction (CPDP). First, we categorize the CPDP literature into representative groups based on their methodological characteristics. Then, we discuss dynamic classifier and ensemble selection techniques and motivate their applicability to CPDP scenarios.
\\

\noindent \textbf{CPDP Groups.} Amasaki, Aman, and Yokogawa~\cite{amasaki2020exploratory} categorized CPDP approaches implemented in the CrossPare library~\cite{herbold2018comparative} into four groups: Data Transformation, Instance Weighting, Ensemble Learning, and Feature Selection. Based on more recent studies, we extend this taxonomy by adding a fifth group related to Advanced Machine Learning methods.

Data transformation methods attempt to reduce distribution shifts by normalizing or scaling metrics across projects~\cite{watanabe2008adapting,cruz2009towards}. More recent approaches extend this idea by learning richer representations through semantic and syntactic encoding of software artifacts, aiming to bridge cross-project gaps at the representation level~\cite{jiang2024syntactic}. Other works select source projects that are closer to the target project according to distance measures~\cite{liu2019two,xia2016hydra}. However, these approaches often require access to the target project distribution and are limited when dealing with entirely new projects or projects with very few instances.

Instance weighting approaches aim to favor training instances that are more similar to the target project. Representative works include relevance filtering using nearest neighbors~\cite{turhan2009relative} and local modeling strategies based on clustering~\cite{menzies2011local}. Although effective in some scenarios, these methods may discard useful data and suffer from high false alarm rates, especially when instance selection is performed at coarse granularity levels~\cite{sinaga_ahmad_abal_2020}. Effort-aware approaches further incorporate inspection cost into the weighting and prioritization process, as demonstrated by recent studies revisiting supervised and unsupervised CPDP methods under effort-aware evaluation settings~\cite{ni2020revisiting}.

Ensemble learning has also been explored in CPDP, with methods combining classifiers trained on different projects or feature subsets~\cite{panichella2014cross,zhang2015empirical}. Nevertheless, static ensembles remain limited by distribution shifts, as only a subset of classifiers is typically competent for predicting a given module, while others may introduce noise.

Feature selection methods focus on improving transferability by removing noisy or redundant metrics~\cite{amasaki2015improving,he2015empirical}. While such approaches simplify the learning problem, they do not explicitly address local competence or instance-level variability across projects.

More recent studies have increasingly focused on advanced machine learning techniques, particularly deep learning and transfer/domain adaptation frameworks. Examples include autoencoder-based models with dynamic adversarial adaptation~\cite{zhang2025cross}, multi-stage cross-project frameworks based on feature representation and knowledge transfer~\cite{zou2025three}, and approaches combining fuzzy embeddings with deep learning models~\cite{azzeh2026cross}. Despite their promising results, these methods generally rely on global feature representations and implicit assumptions about source–target similarity, which may limit their robustness under severe domain mismatch.

\noindent \textbf{Dynamic Classifier and Ensemble Selection.} Dynamic selection (DS) techniques select classifiers based on their estimated competence for each test instance and have demonstrated superior performance over static ensembles in heterogeneous and non-stationary scenarios~\cite{cruz2018dynamic}. These methods assume classifier specialization across different regions of the feature space and dynamically adapt predictions at inference time. DS techniques are typically divided into Dynamic Classifier Selection (DCS), which selects a single classifier, and Dynamic Ensemble Selection (DES), which selects a subset of classifiers per instance. While DCS approaches may be sensitive to noise and class imbalance, DES methods are generally more robust by aggregating multiple competent classifiers, including oracle-based techniques~\cite{ko2008dynamic} and meta-learning approaches such as META-DES~\cite{cruz2015meta,jiao2022dynamic}.

\noindent \textbf{Critical analysis.} Despite the extensive literature on CPDP, Herbold et al.~\cite{herbold2017comparative} showed that existing approaches do not employ dynamic selection mechanisms and generally rely on global modeling assumptions that require information from the target project. Since classifier competence can vary across different regions of the feature space~\cite{cruz2018dynamic}, using a single global model or static ensemble may be inadequate in heterogeneous CPDP scenarios. Dynamic selection addresses this limitation by selecting models at the instance level rather than relying on a single global representation. The proposed Multi-DES framework tackles this challenge by adopting a multi-criteria configuration selection strategy that promotes better generalization across multiple source projects and applies the selected configuration to unseen target projects.

\section{The Proposed Framework: Multi-DES}
\label{sec:MDS-CPDP}

\begin{figure*}[ht]
\centering
\includegraphics[width=18cm]{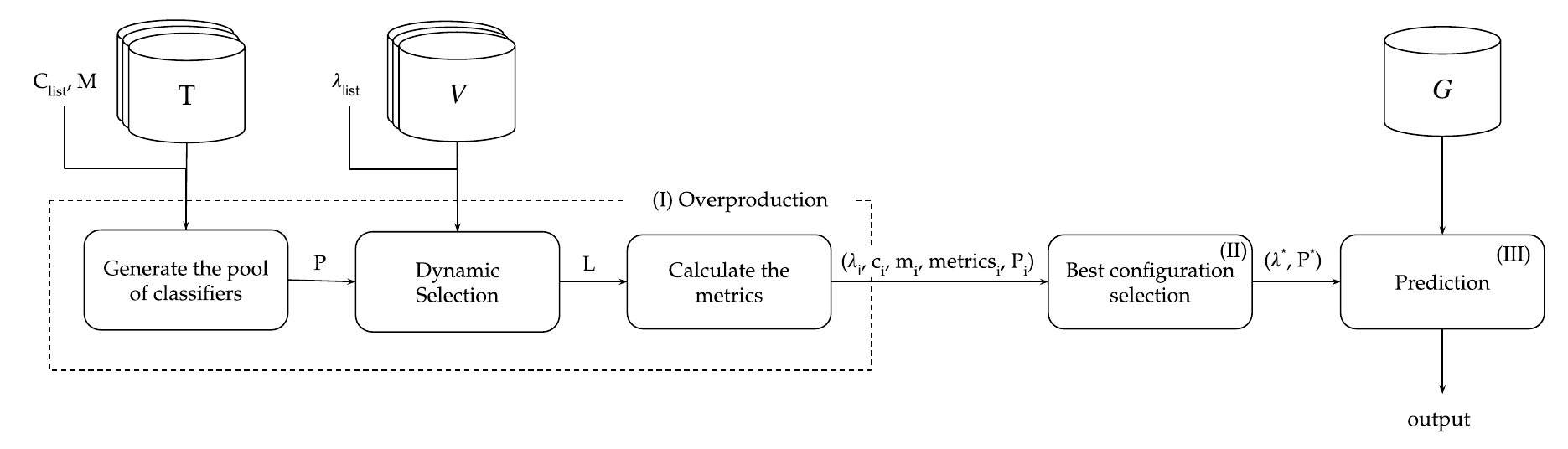}
\caption{Proposed Multi-DES architecture and its three phases: I)~Overproduction, II)~Best configuration selection, and III)~Prediction. $\mathcal{T}$, $\mathcal{V}$, and $\mathcal{G}$ are the training, validation, and test datasets, respectively. $C_{list}$ is a list of learning algorithms, $M$ is the set with different pool sizes used to generate the pool $P$, $\lambda_{list}$ is a list of dynamic selection techniques, P is the pool of classifiers, and L stores the output of the dynamic selection module.}
%\\
%\textbf{Note}: All projects* indicates all projects in \textit{D} except \textit{G} and all other versions of the same project.}
\label{fig:MDS-CPDP_diagram}
\end{figure*}

Figure~\ref{fig:MDS-CPDP_diagram} shows the proposed framework, Multi-stage Dynamic Ensemble Selection (Multi-DES) for Cross-Project Defect Prediction, which aims to select the best model, from a predefined set of options, to predict defects in a project that was not previously used to train the system. Multi-DES is a strict cross-project defect predictor, meaning that it relies exclusively on data from source projects and does not use any information from the target project during training. In particular, the framework exploits a set of source datasets $\mathcal{T} = \{\mathcal{D}_1, \mathcal{D}_2, \ldots, \mathcal{D}_n\}$, where each dataset corresponds to a distinct software project composed of multiple modules, to predict defects in a new target project $\mathcal{G}$ that did not take part in the training process. Following the taxonomy adopted by Herbold et al.~\cite{herbold2017global}, Multi-DES falls into the category of strict cross-project defect prediction.

Multi-DES adopts a multi-stage strategy to optimize the configurations of dynamic selection algorithms using only source projects. This optimization relies on a validation set ($\mathcal{V}$), which, similarly to $\mathcal{T}$, is composed of a set of projects, and aims to mitigate distribution differences between source and target data, one of the main challenges in CPDP systems~\cite{Li22}. As illustrated in Figure~\ref{fig:MDS-CPDP_diagram}, the framework comprises three main phases: I)~overproduction, II)~best configuration selection, and III)~prediction. In the overproduction phase, multiple DS configurations are generated by varying the base learning algorithm, the DS technique, and the pool size. The best configuration is then selected in Phase~II based on its generalization ability across projects. Finally, in the prediction phase, the selected configuration is applied to the target project, dynamically adapting the ensemble to each software module and producing defect predictions for project $\mathcal{G}$.

\subsection{Overproduction}

This phase takes as input the training set $\mathcal{T}$ and the validation set $\mathcal{V}$, together with a set of dynamic selection techniques $\lambda_{list} = \{\lambda_1, \ldots, \lambda_{n_1}\}$, a set of base learning algorithms $C_{list} = \{c_1, \ldots, c_{n_2}\}$, and a set of pool sizes $M = \{m_1, \ldots, m_{n_3}\}$. For each classifier $c \in C_{list}$ and pool size $m \in M$, a classifier pool $P(c,m)$ is generated from $\mathcal{T}$ using Bagging, which promotes diversity through bootstrap sampling. Such diversity is crucial in CPDP scenarios to mitigate cross-project data heterogeneity~\cite{turhan2009relative}.

Each pool is then evaluated on $\mathcal{V}$ using every dynamic selection strategy $\lambda \in \lambda_{list}$. For each instance $v \in \mathcal{V}$, a region of competence $\mathrm{RoC}_{\lambda}(g)$ is defined, and a subset of competent classifiers $P'_{\lambda}(g) \subseteq P(c,m)$ is selected and combined according to $\lambda$ to produce a prediction $l_g$. The predictions obtained for all instances in $\mathcal{V}$ are then used to assess the overall performance of the corresponding dynamic selection configuration, using performance metrics such as F1-score and AUC.

This process evaluates all configurations in the Cartesian product $\lambda_{list} \times C_{list} \times M$, resulting in tuples $(\lambda, c, m, \textit{metrics}, P)$. Exploring this configuration space is crucial for CPDP, as it increases system diversity by varying competence definitions, selection criteria, and pool sizes—factors that are often fixed in prior work~\cite{cruz2018dynamic}. By systematically assessing a wide range of configurations across multiple source projects, Multi-DES identifies configurations that are more likely to generalize under cross-project distribution shifts. Importantly, this phase relies exclusively on source projects, enabling configuration selection without access to the target project $\mathcal{G}$ and allowing defect prediction for entirely new projects.

\subsection{Best configuration selection}
\label{sec:selection}

This phase (Algorithm~\ref{alg:selection}) selects the most competent model based on the results of the experimental configurations generated in the overproduction phase. Each configuration is represented by a tuple $(\lambda, c, m, \textit{metrics}, P)$, where $\lambda$ denotes the dynamic selection algorithm, $P$ is the pool of classifiers trained using the base learning algorithm $c$ with pool size $m$, and $\textit{metrics}$ stores the evaluation results, e.g., F1-score, AUC, and False Alarm.

\begin{algorithm}
\caption{Best configuration selection using Aggregate Rank Minimization (ARM)}
\label{alg:selection}
\begin{algorithmic}[1]
\Require results: list of tuples $(\lambda, c, m, \textit{metrics}, P)$  
    \State $rankings = \varnothing$
    \For{{\bf each} {\it metric} $\in$ $\textit{metrics}$}
        \State $values =$ values-per-metric($\textit{metrics}$.{\it metric})
        \State $rankings[{\it metric}] = \textit{rankdata}(values)$
    \EndFor
    \State $\textit{avg\_ranking} = $ compute-average-rank($rankings$) 
    \State $\textit{low\_rank} = $ select-lower-average($\textit{avg\_ranking}$)
    \State $\lambda^{*}, P^* = $ select-config(results, $\textit{low\_rank}$) \\
    \Return $\lambda^{*}, P^*$
\end{algorithmic}
\end{algorithm}

Different performance metrics are considered in this phase due to the importance of evaluating models from multiple perspectives, as approaches may exhibit varying behavior depending on the criterion adopted~\cite{herbold2017comparative}. Accordingly, we propose an \emph{Aggregate Rank Minimization (ARM)} strategy for configuration selection. ARM operates as follows: for each performance metric, configurations are ranked according to their performance, with the best-performing configuration receiving rank 1, the second-best rank 2, and so forth (lines 2--5). In the presence of ties, a mid-rank (fractional ranking) strategy is adopted, assigning each tied configuration the average of the ranks they would occupy.

After computing rankings for all metrics, ARM calculates the average rank of each configuration across metrics (line 6). The configuration with the lowest average rank (line 7) is selected, as it represents the configuration with the best overall trade-off among the considered metrics and, consequently, the highest expected generalization ability (line 8).

By aggregating rankings across multiple evaluation criteria, ARM balances competing performance objectives and reduces the risk of overfitting to a single metric. As a result, it increases the likelihood of selecting configurations that perform robustly when transferred to unseen target projects.

\subsection{Prediction}

 The \emph{Best configuration selection} phase returns an optimal configuration defined as a pair
$(\lambda^{*}, P^{*})$, where $\lambda^{*}$ denotes the selected dynamic selection strategy and
$P^{*}$ the corresponding pool of classifiers. This configuration is then applied to predict
defects in the target project $\mathcal{G}$.

Formally, for each instance $g \in \mathcal{G}$, a region of competence $\mathrm{RoC}(g)$ is computed based on the validation set $\mathcal{V}$. The region of competence is defined as a subset of instances in $\mathcal{V}$ that are considered relevant for assessing the competence of classifiers with respect to $g$, according to the dynamic selection strategy in use. Given $\mathrm{RoC}(g)$, a subset of classifiers $P'(g) \subseteq P^{*}$ is selected such that each classifier in $P'(g)$ demonstrates adequate competence over $\mathrm{RoC}(g)$. The dynamic selection strategy $\lambda^{*}$ is then applied to select or combine classifiers from $P'(g)$ in order to produce the predicted label $r_g \in \{0,1\}$, indicating the absence or presence of a defect in instance $g$.

\section{Experimental Protocol}
\label{sec:protocol}

\noindent\textbf{Datasets.} Experiments were conducted on four widely used public defect prediction datasets: (1) PROMISE~\cite{jureczko2010towards}, comprising 62 product versions from 31 projects; (2) RELINK~\cite{wu2011relink}, containing defect data from three projects; (3) NASA (MDP), a pre-processed version of the Metrics Data Program with 12 projects~\cite{shepperd2013data}; and (4) AEEEM~\cite{dambros2010}, which includes five Java products from different projects. Each dataset provides a distinct set of software metrics used as features. To handle scale differences, Z-score normalization is applied using statistics computed from the training set and then reused to normalize test instances.

A leave-one-project-out evaluation protocol is adopted, where one project is used as the target and the remaining projects for training. When a project is selected as the target, all its versions are excluded from the training set. The process is repeated for all projects, and the average and standard deviation of the results are reported.

The approaches are evaluated using three performance metrics—F1-score, Area Under the ROC Curve (AUC), and False Alarm—and statistical significance is assessed using the Wilcoxon Signed-Rank test.

%\subsection{Classifiers and Dynamic Selection Techniques}\label{sec:classifiers-and-ds}

\noindent\textbf{Classifiers and Dynamic Selection Techniques.} Dynamic selection techniques rely on a pool of base classifiers; thus, we employed \textit{Decision Tree} (DT), \textit{Logistic Regression} (LR), \textit{Naive Bayes} (NB), and \textit{Random Forest} (RF), which are widely used in CPDP and dynamic ensemble learning~\cite{herbold2017comparative,cruz2018dynamic}. All classifiers were implemented using scikit-learn\footnote{\url{https://scikit-learn.org/}
 --- version 1.0.2} with default hyperparameters. The classifier pool was generated using Bagging, the standard approach in the dynamic selection literature~\cite{cruz2015meta,cruz2018dynamic}, with pool sizes ranging from 10 to 100 in steps of 10.

Eight state-of-the-art dynamic selection techniques were considered: \textit{K-Nearest Oracles Union} (KNU) and \textit{K-Nearest Oracles Eliminate} (KNE)~\cite{ko2008dynamic}, \textit{k-Nearest Output Profiles} (KNOP)~\cite{cavalin2013dynamic}, \textit{META-DES}~\cite{cruz2015meta}, \textit{Overall Local Accuracy} (OLA) and \textit{Local Class Accuracy} (LCA)~\cite{woods1997combination}, \textit{Multiple Classifier Behavior} (MCB)~\cite{giacinto2001dynamic}, and \textit{Modified Classifier Rank} (Rank)~\cite{sabourin1993classifier}. These techniques were selected based on citation impact, diversity of selection criteria (oracle, behavior, accuracy, and meta-learning), reported performance, and availability in DESlib\footnote{\url{https://github.com/scikit-learn-contrib/DESlib}
 --- version 0.3.5}~\cite{cruz2020deslib}.

%\subsection{State-of-the-art CPDP approaches}\label{sec:sota-cpdp}

\noindent \textbf{State-of-the-art CPDP approaches.} The proposed framework is compared with four representative CPDP methods—\textit{CamargoCruz09-DT} (CC09-DT)~\cite{cruz2009towards}, \textit{Turhan09-DT} (T09-DT)~\cite{turhan2009relative}, \textit{Menzies11-RF} (M11-RF)~\cite{menzies2011local}, and \textit{Watanabe08-DT} (W08-DT)~\cite{watanabe2008adapting}—which were identified as top-performing techniques in the large-scale comparative study by Herbold et al.~\cite{herbold2017comparative}. The experimental results of these methods were obtained from the publicly available replication package\footnote{\url{https://github.com/sherbold/replication-kit-tse-2017-benchmark}}
. In addition, we compare our framework with the Effort-Aware Supervised Cross-project method~\cite{ni2020revisiting}, using Naive Bayes as the base classifier (EASC-NB), which was proposed after the study by Herbold et al \cite{herbold2017comparative}.

%\textcolor{red}{As to the EASC-NB, we used the results obtained from replicating the method according to the guidelines and original configurations introduced by Ni et al. \cite{ni2020revisiting}. It is also worth highlighting that the EASC-NB has a limitation concerning the types of data to which the method can be applied. This method requires the dataset to provide information on the number of lines of code (LOC) that a certain software module has. Such information is an independent variable in the dataset and refers to and/or is related to the effort/time required to inspect a specific instance (software module).}

%\subsection{Evaluation}

\section{Experiments}	
\label{sec:experiments}

The experimental analysis evaluates the proposed Multi-DES framework from three complementary perspectives. First, it examines whether the multi-stage design effectively exploits the diversity generated in the overproduction phase by analyzing the contribution of different configurations across projects. Second, it assesses the effectiveness of the proposed Ranking strategy for configuration selection, focusing on its ability to identify models with higher generalization capability through the joint consideration of multiple evaluation metrics. Finally, Multi-DES is compared with state-of-the-art CPDP approaches to analyze its relative performance across datasets and evaluation metrics.

\begin{figure*}[t]
\centering
\includegraphics[width=0.32\textwidth]{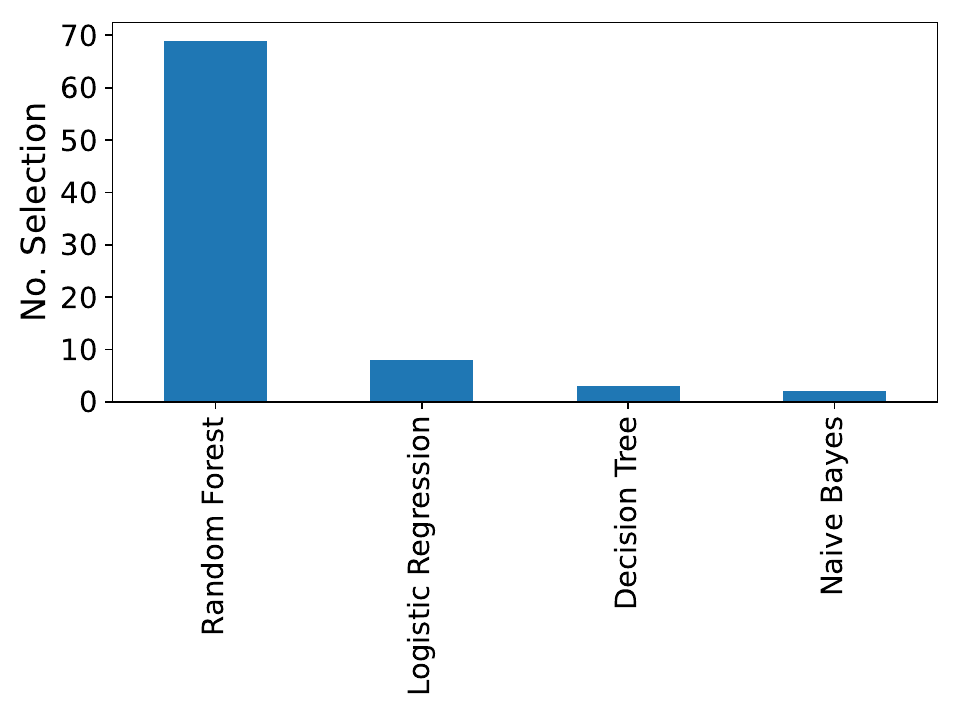}
\includegraphics[width=0.32\textwidth]{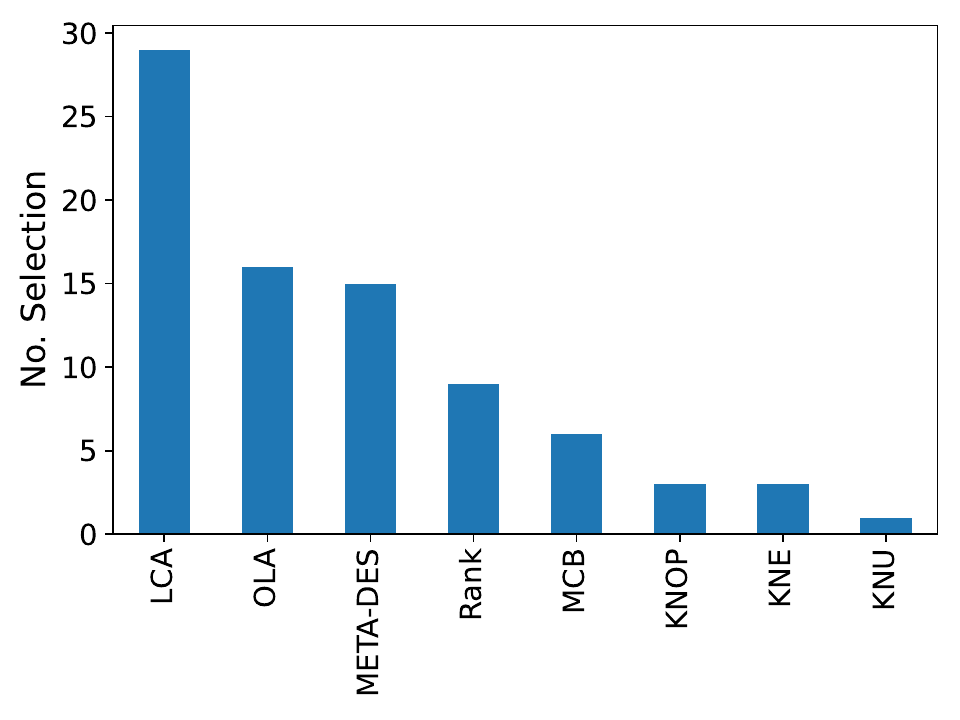}
\includegraphics[width=0.32\textwidth]{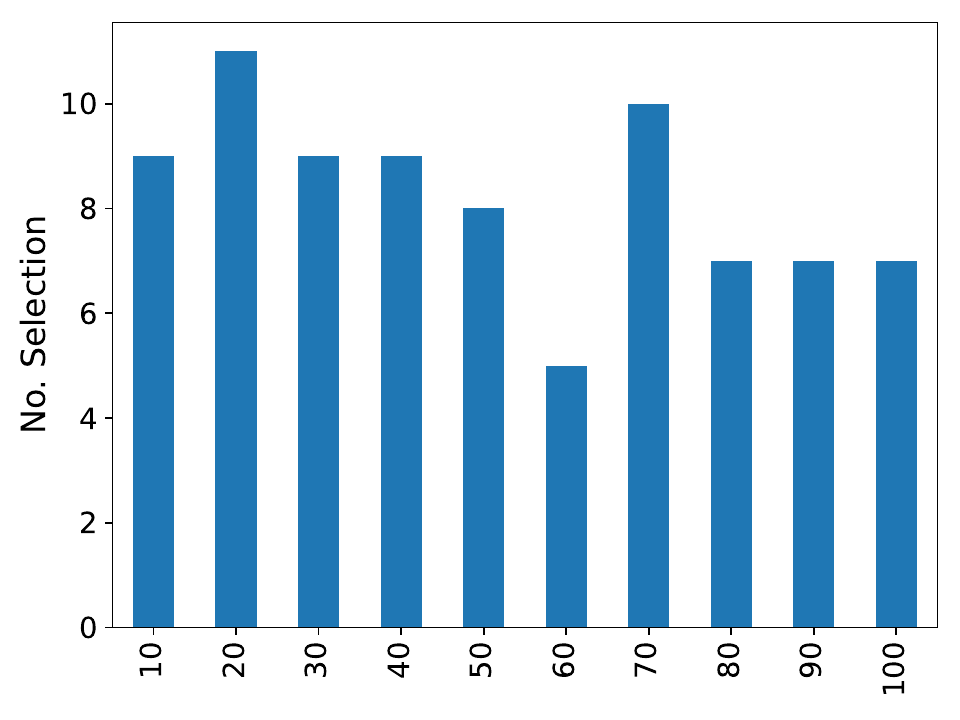}
\caption{Configuration selection frequency: (a) base classifier, (b) dynamic selection method, and (c) pool size.}
\label{fig:selectionDistribution}
\end{figure*}

\subsection{Configuration analysis}
\label{sec:parameters-analysis}

In the Overproduction phase, four base learning algorithms ($C_{list}$), eight dynamic selection techniques ($\lambda_{list}$), and then pool sizes ($M$) are evaluated, generating a total of 320 ($4 \times 8 \times 10$) configurations. After that, only one configuration is selected in the second phase, ``Best configuration selection'', per project under evaluation. These configurations compose diverse classification systems that create different decision spaces. 

Figure~\ref{fig:selectionDistribution} shows the number of times each configuration is selected. Since the leave-one-project-out procedure is employed, 82 configurations are selected, i.e., one per project. For the base classifiers (Figure~\ref{fig:selectionDistribution}(a)), Random Forest deserves a remark, being selected 69 out of 82 times. However, other base classifiers were the best option depending on the project under evaluation. Regarding the dynamic selection technique (Figure~\ref{fig:selectionDistribution}(b)), LCA, OLA, and META-DES reached the top three methods, being selected 29, 16, and 15 times, respectively. Concerning the pool size (Figure~\ref{fig:selectionDistribution}(c)), no clear preference can be observed, as the distribution is nearly uniform. This behavior is consistent with previous in-depth analyses of dynamic ensemble selection frameworks, which have shown that different pool sizes often lead to comparable performance, with observed variations largely attributable to randomness rather than systematic effects.

Important to highlight that all the possible configurations of base classifier, dynamic selection method, and pool size were selected at least once. In particular, there was no clear better option for the dynamic selection method and the pool size. These results show the relevance of having a procedure to search for the best configuration instead of adopting a static classifier, since the best configuration differs from task to task.

\subsection{Selection approach evaluation}
\label{sec:selection-approach-evaluation}

Four strategies were evaluated to select the best configuration in Phase II of the Multi-DES framework: three single-metric strategies optimizing F1-score, AUC, or False Alarm, and the proposed Aggregate Rank Minimization (ARM) strategy, which combines multiple metrics to favor configurations with better generalization capability.

The results show that ARM achieves superior or comparable performance to single-metric strategies across most datasets and evaluation metrics. Pairwise Wilcoxon signed-rank tests indicate that ARM outperforms or matches single-metric selection in the majority of cases ($p < 0.05$), particularly when compared with False Alarm--based selection, where ARM consistently identifies more effective configurations.

These findings reveal a key limitation of single-metric selection, as optimizing a configuration for a single criterion may lead to overfitting and reduced robustness on unseen projects, reinforcing the need for multi-perspective evaluation in CPDP~\cite{herbold2017comparative}. To further contextualize its effectiveness, ARM is compared with an upper-bound scenario (\emph{Truth}), which represents an idealized retrospective selection based on test performance. Although infeasible in practice, this comparison shows that ARM often closely approximates the upper bound, especially for AUC, while larger gaps in some cases (e.g., False Alarm on RELINK) indicate opportunities for improvement.

Overall, the results confirm that no single configuration is optimal across all projects and highlight the effectiveness of the proposed multi-stage selection scheme, supporting the adoption of ARM as the default selection strategy in subsequent analyses.

\subsection{Multi-DES versus state-of-the-art}
\label{sec:muli-des-sota}

Table~\ref{tab:cpdp-compare-unified} presents the performance of Multi-DES and state-of-the-art CPDP methods across four datasets. The best results per dataset are highlighted in bold, and the second best are underlined.

\begin{table*}[t]
\centering
\caption{Performance (mean$\pm$variance) of Multi-DES and CPDP methods across datasets.
The last row of each metric reports the p-values of the Wilcoxon signed-rank test comparing
Multi-DES against each baseline across all datasets. Best values per dataset are in bold and
second best are underlined.}
\label{tab:cpdp-compare-unified}
\begin{tabular}{llllllll}
\hline
Metric & Dataset & CamargoCruz09-DT & EASC-NB & Menzies11-RF & Turhan09-DT & Watanabe08-DT & Multi-DES \\
\hline

\multirow{5}{*}{F1-score $\uparrow$}
& AEEEM   & 0.312$\pm$0.007 & \underline{0.372$\pm$0.010} & 0.269$\pm$0.008 & 0.271$\pm$0.004 & 0.301$\pm$0.000 & \textbf{0.385$\pm$0.009} \\
& NASA    & 0.089$\pm$0.007 & \underline{0.150$\pm$0.011} & 0.142$\pm$0.003 & \textbf{0.161$\pm$0.007} & 0.109$\pm$0.002 & 0.140$\pm$0.006 \\
& PROMISE & \underline{0.367$\pm$0.023} & 0.270$\pm$0.020 & 0.328$\pm$0.016 & 0.359$\pm$0.029 & \textbf{0.368$\pm$0.014} & 0.316$\pm$0.022 \\
& RELINK  & 0.543$\pm$0.003 & 0.242$\pm$0.118 & \underline{0.555$\pm$0.009} & 0.528$\pm$0.057 & 0.489$\pm$0.022 & \textbf{0.560$\pm$0.015} \\ \hline
& Wilcoxon $p$ & 0.2382 & \textbf{0.0237} & 0.7939 & 0.3421 & 0.1763 & -- \\
\hline

\multirow{5}{*}{AUC $\uparrow$}
& AEEEM   & 0.604$\pm$0.005 & \underline{0.692$\pm$0.006} & 0.578$\pm$0.001 & 0.530$\pm$0.000 & 0.588$\pm$0.004 & \textbf{0.755$\pm$0.002} \\
& NASA    & 0.697$\pm$0.013 & 0.666$\pm$0.006 & 0.541$\pm$0.001 & 0.622$\pm$0.004 & \underline{0.669$\pm$0.009} & \textbf{0.737$\pm$0.008} \\
& PROMISE & 0.582$\pm$0.010 & \textbf{0.713$\pm$0.013} & 0.574$\pm$0.005 & 0.588$\pm$0.012 & 0.593$\pm$0.008 & \underline{0.709$\pm$0.012} \\
& RELINK  & 0.648$\pm$0.001 & \textbf{0.748$\pm$0.006} & 0.653$\pm$0.001 & 0.635$\pm$0.006 & 0.602$\pm$0.011 & \underline{0.734$\pm$0.010} \\ \hline
& Wilcoxon $p$ & \textbf{0.000}$^*$ & 0.5494 & \textbf{0.000}$^*$ & \textbf{0.000}$^*$ & \textbf{0.000}$^*$ & -- \\
\hline

\multirow{5}{*}{False Alarm $\downarrow$}
& AEEEM   & \underline{0.063$\pm$0.001} & 0.174$\pm$0.021 & 0.047$\pm$0.002 & 0.131$\pm$0.009 & 0.109$\pm$0.002 & \textbf{0.035$\pm$0.000} \\
& NASA    & \textbf{0.013$\pm$0.000} & 0.021$\pm$0.000 & 0.034$\pm$0.001 & 0.052$\pm$0.002 & 0.019$\pm$0.000 & \underline{0.019$\pm$0.000} \\
& PROMISE & 0.203$\pm$0.012 & \textbf{0.063$\pm$0.003} & 0.171$\pm$0.008 & 0.178$\pm$0.010 & 0.262$\pm$0.019 & \underline{0.097$\pm$0.006} \\
& RELINK  & 0.212$\pm$0.006 & 0.170$\pm$0.058 & 0.211$\pm$0.002 & \textbf{0.167$\pm$0.015} & \underline{0.168$\pm$0.004} & 0.190$\pm$0.005 \\ \hline
& Wilcoxon $p$ & \textbf{0.000}$^*$ & 0.0023 & \textbf{0.000}$^*$ & \textbf{0.000}$^*$ & \textbf{0.000}$^*$ & -- \\
\hline
\end{tabular}

\flushleft\footnotesize{
$\uparrow$ indicates higher is better; $\downarrow$ indicates lower is better.
Bold p-values indicate statistical significance at $\alpha = 0.05$.
$^*$ indicates $p < 10^{-5}$.
}
\end{table*}

For the F1-score, Multi-DES achieved the best performance on the AEEEM and RELINK datasets. Regarding AUC, Multi-DES obtained the best results on AEEEM and NASA, and second-best results on PROMISE and RELINK. In the latter two datasets, the differences between Multi-DES and EASC are small (around 1\%), while in AEEEM and NASA, Multi-DES outperformed EASC by approximately 10\%. For the False Alarm metric, Multi-DES achieved the best or second-best performance in most datasets, indicating robust behavior across different scenarios. Notably, on the AEEEM dataset, Multi-DES achieved the best results for all metrics and was never the worst-performing method.

The Wilcoxon signed-rank test results reported in the last row of each metric block in Table~\ref{tab:cpdp-compare-unified} indicate that Multi-DES is statistically superior in most pairwise comparisons, particularly for the AUC and False Alarm metrics.

Table~\ref{tab:win-tie-loss-agg} summarizes the aggregated win--tie--loss counts comparing Multi-DES against each baseline across all datasets. For AUC, Multi-DES clearly dominates the competing methods, while for False Alarm, it also achieves a large number of wins against most baselines. Although the F1-score presents a more balanced scenario, Multi-DES remains competitive and achieves comparable or superior performance to the literature methods.

\begin{table*}[t]
\centering
\caption{Aggregated win--tie--loss results comparing Multi-DES against literature CPDP methods across all datasets.}
\label{tab:win-tie-loss-agg}
\begin{tabular}{lccccc}
\hline
Metric & CamargoCruz09-DT & EASC-NB & Menzies11-RF & Turhan09-DT & Watanabe08-DT \\
\hline
F1-score    & 40--2--40 & 43--4--35 & 41--4--37 & 33--2--47 & 37--2--43 \\
AUC         & 70--1--11 & 42--1--39 & 76--0--6  & 77--0--5  & 72--0--10 \\
False Alarm & 61--6--15 & 22--13--47 & 65--9--8  & 69--9--4  & 69--5--8 \\
\hline
\end{tabular}
\end{table*}

Considering that the maximum number of wins per metric is 82, resulting in 246 wins across all three metrics, Multi-DES achieved more than 70\% of the wins against CamargoCruz09-DT, Menzies11-RF, Turhan09-DT, and Watanabe08-DT, and a comparable but slightly higher number of wins than EASC-NB.

It is worth noting that Multi-DES builds and selects models exclusively using training data, unlike several CPDP approaches that rely on information extracted from the target project for model selection or data transformation, which may lead to data leakage. Despite this restriction, Multi-DES achieves superior or equivalent results without leveraging any information from the test project. Moreover, due to its local and instance-based dynamic selection strategy, Multi-DES does not make assumptions about the target project distribution, making it particularly suitable for defect prediction in new projects with limited available data.

\section{Conclusion}
\label{sec:conclusion}

Software defect prediction remains challenging due to limited labeled data, motivating the use of Cross-Project Defect Prediction (CPDP) to leverage external projects. This paper proposes Multi-DES, a multi-stage dynamic ensemble selection framework that selects model configurations using only source projects and adapts predictions at the instance level. In particular, we introduce the proposed \emph{Aggregate Rank Minimization (ARM)} strategy for configuration selection, which improves generalization by jointly considering multiple performance metrics. The experimental results show that ARM improves generalization and that Multi-DES achieves superior or comparable performance to state-of-the-art CPDP methods across most datasets and metrics, particularly under distribution shifts, despite a single exception on the PROMISE dataset for False Alarm. The framework requires no information from the target project, enabling its use on completely new projects. Validity threats were addressed through a leave-one-project-out protocol on 82 projects, widely used evaluation metrics, and appropriate statistical tests. Future work includes reducing computational cost via configuration prediction, extending Multi-DES to online scenarios, and exploring its applicability to other defect prediction settings.

\bibliographystyle{IEEEtran}
\bibliography{bib}

\end{document}